\documentclass[12pt]{article}
\usepackage[round]{natbib}
\usepackage{times}

\newenvironment{sciabstract}{\begin{quote} \bf}
{\end{quote}}

\usepackage[utf8]{inputenc}
\usepackage[T1]{fontenc}
\usepackage{tcolorbox}
\usepackage{subcaption}
\usepackage{soul}
\usepackage{booktabs,boldline}
\usepackage{fontawesome}
\usepackage{url}
\usepackage{longtable}

\usepackage{multirow}
\usepackage{array}
\newcommand{\PreserveBackslash}[1]{\let\temp=\\#1\let\\=\temp}
\newcolumntype{P}[1]{>{\raggedright\arraybackslash}p{#1}}

\definecolor{darkgray}{rgb}{0.33,0.33,0.33}
\newcommand{\checkbox}{\textcolor{darkgray}{\scalebox{0.85}{\faCheckSquareO}~}}

\newcommand{\rev}[1]{\textcolor{black}{#1}}

\let\oldFootnote\footnote
\newcommand\nextToken\relax

\renewcommand\footnote[1]{\oldFootnote{#1}\futurelet\nextToken\isFootnote}

\newcommand\blfootnote[1]{\begingroup
  \renewcommand\thefootnote{}\footnote{#1}\addtocounter{footnote}{-1}\endgroup
}

\newcommand\isFootnote{\ifx\footnote\nextToken\textsuperscript{,}\fi}

\title{Demystifying Misconceptions in Social Bots Research}

\author
{Stefano Cresci,$^{1,*}$ Kai-Cheng Yang,$^{2}$ Angelo Spognardi,$^{3}$ Roberto Di Pietro,$^{4}$\\Filippo Menczer,$^{5}$ Marinella Petrocchi$^{1}$\\
\\
\normalsize{$^{1}$IIT-CNR, Italy}\\
\normalsize{$^{2}$Northeastern University, USA}\\
\normalsize{$^{3}$``La Sapienza'' University of Rome, Italy}\\
\normalsize{$^{4}$KAUST, Saudi Arabia}\\
\normalsize{$^{5}$Indiana University, USA}\\
\\
}

\date{}

\begin{document} 

\baselineskip18pt
\maketitle 

\begin{sciabstract}
Research on social bots aims at advancing knowledge and providing solutions to one of the most debated forms of online manipulation. Yet, social bot research is plagued by widespread biases, hyped results, and misconceptions that set the stage for ambiguities, unrealistic expectations, and seemingly irreconcilable findings. Overcoming such issues is instrumental towards ensuring reliable solutions and reaffirming the validity of the scientific method.
Here, we discuss a broad set of consequential methodological and conceptual issues that affect current social bots research, illustrating each with examples drawn from recent studies. More importantly, we demystify common misconceptions, addressing fundamental points on how social bots research is discussed. Our analysis surfaces the need to discuss research about online disinformation and manipulation in a rigorous, unbiased, and responsible way. This article bolsters such effort by identifying and refuting common fallacious arguments used by both proponents and opponents of social bots research, as well as providing directions toward sound methodologies for future research.

\blfootnote{$^*$ Corresponding author: stefano.cresci@iit.cnr.it}

\end{sciabstract}

\pagebreak

\section{Introduction}
\label{sec:intro}
Human decision-making processes depend on the availability of high-quality information and a healthy society requires a shared understanding of issues and values. Misinformation and disinformation erode trust and emphasize divisions. Moreover, the presence of divergent or incompatible beliefs can hinder reaching consensus and taking effective collective action. The repercussions could be so severe as to delay responses to a deadly pandemic~\citep{covaxxy-misinfo} or endanger democratic processes~\citep{pennycook2021shifting}. 

The science of misinformation seeks solutions to these problems~\citep{lazer2018science}. However, this paramount endeavor can itself incur the same problems that it aims to overcome \citep{west2021misinformation,altay2023misinformation}. For example, scientific articles and publishers engage in a fierce competition for attention, much like mainstream news outlets. As a consequence, sensationalist claims and hyped results are sometimes used as shortcuts to publication and scientific recognition~\citep{west2021misinformation}. Confirmation bias is also making its way into science, in the form of citation bias: the preference for citing articles that support one's results over those that challenge them~\citep{west2021misinformation}. And again, over-generalizations of scientific results and poor understanding of methodological and conceptual limitations give rise to multiple misconceptions about misinformation~\citep{altay2023misinformation}. These and other problems currently undermine the efficacy and credibility of our research efforts. Therefore, for the science of misinformation to benefit our society, we must first solve the problems within the science itself.

The present study concerns one of the many forms of online manipulation: social bots. Among the many diverse existing definitions~\citep{gorwa2020unpacking}, here we define social bots as social media accounts that are totally or partially automated. Hereafter, we use the terms ``social bot'' and ``bot'' interchangeably, always in line with the aforementioned definition.

Due to their automation, simplicity, and low operating cost, social bots can be easily used as expendable tools for spreading problematic content. Understanding the role and activity of bots in large-scale manipulation campaigns is important, as it can inform strategies to curb mis- and disinformation~\citep{shao2018spread}. For this reason, social bots have attracted considerable scholarly and media attention~\citep{allem2018could,assenmacher2020demystifying,gonzalez2021bots,chen2021neutral}. A recent example is the public dispute over Twitter’s bot count preceding Elon Musk’s acquisition~\citep{varol2022should}. However, despite many years of research, the science of social bots is replete with the same problems that also plague the science of misinformation. These originate from uncorrected biases in how scientific results are cited and discussed, and a wide array of methodological and conceptual issues that set the stage for ambiguities, misunderstandings, unrealistic expectations, as well as contrasting and seemingly irreconcilable findings~\citep{rauchfleisch2020false,hays2023simplistic}.

Here, we develop a critical and theoretically grounded perspective on key methodological and conceptual challenges in social bots research. To this end, we discuss a large set of consequential issues for the field, based on their recurrence, impact, and potential for misunderstanding. In discussing each issue, we refer to a small number of recent studies that serve as illustrative examples of the broader problems. Therefore, while this article does not aim to be exhaustive in terms of the analyzed literature, the range of issues addressed here is extensive and comprehensive---as reflected in Table~\ref{tab:checklist}. Our critical analysis thus serves a twofold goal. On the one hand, we highlight and revise methodological biases and conceptual issues in social bots research. On the other hand, demystifying common misconceptions allows us to address fundamental issues on how science is produced and discussed~\citep{west2021misinformation}. Overcoming such issues is instrumental in ensuring the credibility of science.

 \section{Methodological issues}
\label{sec:issues-method}
In this section, we focus on issues that arise from how social bots research is conducted and evaluated, covering aspects such as data selection, model training, and the dissemination of results.

Social bot detection is a challenging task in the realm of online safety and cybersecurity~\citep{cresci2020decade}. In practice, it often involves the use of machine learning algorithms in a binary classification setting, with the goal of distinguishing human-operated accounts from automated ones---the social bots. The machine learning models are trained on features extracted from user profiles, behaviors, network interactions, and linguistic patterns acquired from user posts, with the goal of detecting subtle differences between authentic users and automated entities. Methodological issues in social bots research include a mix of typical machine learning pitfalls and unique issues inherent in the evolving nature of online social ecosystems, which we exemplify in the following sections.

\subsection*{Model drift}
Model drift arises when the features a model relies on differ between training and prediction data---whether due to gradual shifts over time or inherent dataset differences. It can erode accuracy or falsely suggest strong performance by exploiting training-specific patterns absent at prediction time. In bot detection, the problem may surface when knowledge of some peculiar characteristic of the accounts is known beforehand and exploited at training time, allowing for near-perfect detection performance. For example, each botnet typically exhibits some peculiar characteristics resulting from the bot creation process, their goals, or any other shared characteristic~\citep{zhang2016rise,mazza2022investigating}. These peculiar characteristics might set the bots apart from the average behavior of human-operated accounts~\citep{cresci2020emergent}. However, typically such characteristics are known only for botnets that have already been identified and are instead unknown for still undetected botnets. Here we recall that a classifier's detection performance on a held-out portion of known data serves merely as a proxy for its real-world detection performance in-the-wild. In fact, the end goal of any bot detector is to accurately detect unknown bots from a set of never-seen-before data. However, if a bot detector exploits known characteristics at training time for classifying instances of known bots, it might learn correlations that do not generalize to the unknown bots that the classifier will be tasked to detect later on. 

A specific instance of model drift in bot detection is when a detector exploits knowledge of how the accounts were collected---be them the malicious bots or the genuine ones---as a means to predict their class (i.e, automated or otherwise). The correlation between the information on how certain accounts were collected and their class is spurious with respect to the task of bot detection, which makes features based on such information dangerously misleading. Consequently, a detector exploiting these features would achieve excellent performance on known data but would exhibit poor capacity to generalize to new data with different characteristics. Leveraging features used to select and label the data can also be framed as a feature leakage issue, since those features would unfairly advantage the detector. 
Such a situation can lead to the proposal of ostensibly superior models, which, however, represent a regression rather than an advancement in addressing the bot detection problem. In addition, the same models could also be exploited to unduly undermine the performance of other models. For example, some scholars trained simple but unrealistic bot detectors using features tightly bound to the training dataset, such as whether an account was verified by Twitter---which was among the features used to select and label the data---to exaggerate the limitations of more general state-of-the-art models~\citep{gallwitz2022investigating,hays2023simplistic}.

\subsection*{Unfair evaluation}
Other challenges arise from the multitude of disparate definitions, detectors, and benchmark datasets~\citep{cresci2020decade}. For example, a critical issue in evaluating bot detectors is the possible inconsistency between the bot definition assumed by the detector and that used by the evaluator. This inconsistency can lead to biased and unfair evaluations, as it is conceptually flawed to evaluate a detector using a different definition than that employed by the detector itself. A tool trained to identify a specific type of bot should be assessed based on its ability to detect that specific type, not alternative ones for which it was not designed. This misalignment in definitions can result in evaluations that misrepresent the detector's true performance and capabilities. Therefore, while it is perfectly acceptable to discuss---and even criticize---the social bot definition used by a detector, evaluators must ensure that the definition used to evaluate a detector aligns with the detector's, or be very clear about the differences and their implications towards the results of the evaluation. However, there exists an important trade-off between the need for consistency in using definitions and the necessity of assessing the detector's generalizability, which requires testing it against bots with different characteristics~\citep{cresci2020decade}. Striking this balance depends on multiple factors, including the practical context in which the detector is used and whether the detector is specialized towards specific types of bots or general-purpose.

In addition, the existence of many bot detectors creates an environment where authors introducing new detectors can selectively engage in favorable comparisons. The issue is exacerbated by the difficulty of delineating a clear state-of-the-art in bot detection, as the sheer volume of existing models makes it arduous to discern the most effective ones. The lingering uncertainty around the performance of even established bot detectors~\citep{rauchfleisch2020false} makes it possible to cherry-pick competitors and evaluation scenarios, allowing proponents of a novel detector to demonstrate its superiority only against a small subset of detectors and datasets, conveniently omitting those against which it may not perform as well. 

The evaluation landscape is further complicated by the seemingly excellent performance achieved by models that overfit to specific evaluation datasets.\footnote{We refer to overfitting when a model depends too much on irrelevant features of the training data, with the result that it performs well on the training data but generalizes poorly to unseen data~\citep{bramer2007avoiding}.} The issue arises from the inappropriate comparison of a general-purpose model with an extremely specialized---potentially overfitted---one. As an example, Hays et al.~\citep{hays2023simplistic} selected some datasets to train trivial classifiers with a very small number of features. Then, they compared the trivial classifiers to more complex, general, state-of-the-art models, on the same dataset where the trivial models had been trained, suggesting that very few features are sufficient to identify bots. While this approach can highlight biases in certain datasets, the comparison itself is unfair and should not be used to undermine the utility of general-purpose detectors or to criticize their performance.

Attention should also be devoted to the data used for the evaluation. As an example of problematic use of evaluation data, Gallwitz \& Kreil~\citep{gallwitz2022investigating} challenged the accuracy of a widely-used bot detector based on its results on a small dataset of public figures’ accounts. However, these accounts---typically managed by social media teams---do not represent the broader platform user base, possibly leading to under- or overestimation of the detector’s performance. While useful in a specific context, such evaluation does not justify generalized conclusions about the detector’s effectiveness. More broadly, since bot detection methods---whether based on machine learning or on human annotation---have inherent error margins, accuracy results can be manipulated by carefully choosing test examples.

In light of the pervasive challenges in evaluating results about social bots, a common recommendation is to always manually check a subset of the data after classification~\citep{DBLP:journals/jocss/YangFM22}, as this often allows detecting outright misclassifications and possible underlying problems. Moreover, evaluating detectors should not only involve rigorous testing during training but also continuous validation when using pre-trained models, irrespective of their established reputation.

\subsection*{Cherry-picking}
The issue of cherry-picking extends beyond unfair comparisons. A concerning trend involves selectively including, excluding, or misrepresenting prior literature to propose a narrative that aligns favorably with one's own findings. Recently, this has been done to cast seemingly new criticism against certain bot detectors~\citep{hays2023simplistic}. However, the impression of novelty in such criticism can only be made by omitting a significant amount of literature~\citep{varol2017online,Yang2019botometer,yang2020scalable,sayyadiharikandeh2020detection,cresci2020decade}. An opposite---but equally misleading---practice involves repeatedly citing one's own unpublished results, which could give a false impression of prior research supporting one's claims~\citep{gallwitz2022investigating}. Selective referencing can not only distort the perceived reliability and novelty of a work, but also skew the representation of the state-of-the-art. While citing unpublished work is not always bad \textit{per se}, by self-citing under different forms multiple unvetted claims about the results of a single analysis, authors may create an illusion of authority. And by omitting works that have already made certain contributions or conclusions, authors may create an illusion of originality, potentially overshadowing a substantial body of pre-existing research that has contributed comparable insights. This not only undermines the integrity of the academic discourse but also detracts from the collective acknowledgment and recognition owed to the broader community of scholars who have previously advanced the field.

Addressing the cherry-picking issue necessitates a reliance on expert reviewers deeply versed in the nuances of the field, capable of discerning the strategic omission of relevant literature, definitions, methods, or data. Paradoxically, the escalating trend of publications~\citep{haghani2022trends}, particularly in hot topics like bot detection~\citep{cresci2020decade}, poses a formidable challenge, as the growing demand for reviewers outpaces the available pool of experts. This discrepancy surfaces a crucial tension in maintaining the quality and rigor of peer review processes within rapidly evolving research landscapes~\citep{vanmore}. The field is now confronted not only with the difficulty of establishing a definitive benchmark, but also with the challenge of navigating through a heterogeneous literature landscape, where discerning genuine advancements becomes a complex task amid the noise produced by contributions of varying quality.

\subsection*{Straw-man methodology}
The straw-man fallacy consists of misrepresenting someone else's research and then criticizing the misrepresentation instead of the original research. A common manifestation is the claim that bot detection is exclusively a supervised machine learning task. This wrong assumption could perhaps be explained by the fact that supervised machine learning is the traditional way in which the task was tackled, and by the multitude of existing supervised bot detectors~\citep{cresci2020decade}. Despite their prevalence, however, a robust body of literature highlighted that supervised detectors suffer from lack of generalizability and transferability~\citep{cresci2020decade,echeverri2018lobo,yang2020scalable,sayyadiharikandeh2020detection,rauchfleisch2020false,dimitriadis2021social}. A variety of unsupervised and semi-supervised detectors were proposed as possible solutions to these problems, considering groups instead of single accounts and studying features like the temporal patterns of their actions or the structures of their graph representations~\citep{cresci2020decade}. Nonetheless, some studies specifically focus on evaluating the generalization capabilities of bot detectors, but only consider supervised detectors with known generalization deficiencies. The unsurprising result is that the considered detectors fail to generalize, which is used to criticize the whole field---a textbook example of circular reasoning. For instance, Hays et al.~\citep{hays2023simplistic} criticize the generalization capabilities of all bot detectors, despite having experimented only with a narrow set of supervised methods. Similarly, Gallwitz \& Kreil~\citep{gallwitz2022investigating} conclude that «the field of social bot research is fundamentally flawed», despite having investigated only a tiny fraction of the research based on one single bot detector. The issue lies in the broad conclusions of such studies, which are unsupported by their methodology and results. Neither of the two aforementioned works made distinctions between the different types of bot detectors proposed to date. A further example is the claim that a systematic evaluation of bot predictions in real-world scenarios have never been done~\citep{gallwitz2022investigating}. The accusation points to a lack of manual validation and publicly available datasets. However, the majority of existing bot detectors have indeed been evaluated on training and test datasets, with many of these datasets being publicly available. While we acknowledge that such evaluations may have inherent limitations, and that it is important to manually validate bot labels and to share this validation data, the generic criticism that no systematic evaluation has been done before is unwarranted and potentially misleading.

Therefore, future work should avoid oversimplifying social bot detection as a solely supervised task. Instead, works that propose new detectors, that use already existing ones, or that discuss the state-of-the-art, should acknowledge the multitude of approaches to the task, with their strengths and weaknesses. Then, the type of evaluation, as well as the type of bot detector to develop, use, or discuss in a given study should be chosen so as to be adequate and consistent with the objective of the study.

\subsection*{Data biases}
In machine learning and data science, high-quality data is the linchpin for robust model development and insightful analyses~\citep{halevy2009unreasonable}. As mentioned above, numerous datasets of bot and human accounts have been published over the years. On the one hand, this made it easier to thoroughly test the performance of new bot detectors. On the other hand, it introduced the risk of cherry-picking favorable datasets, potentially distorting reported performances and introducing bias. Furthermore, the temporal heterogeneity stemming from the availability of datasets spanning varying time periods---ranging from recent to decade-old---can pose additional challenges when detectors are trained on outdated data, undermining the relevance of their performance in contemporary settings. When considering the representativeness of published benchmark datasets with respect to the bots that currently inhabit online platforms, the readers should be aware that the accounts included in a dataset are likely a few years older than the publication date of the dataset itself. Hence, the publication date of a dataset should be considered as a generous upper bound of the recency of the accounts therein. Even under this relaxed assumption, in consideration of the rapidly evolving landscape of online harms and the availability of more comprehensive and recent datasets~\citep{feng2021twibot}, we risk relying on datasets that are no longer representative of the actual state of the platforms. This is particularly troublesome in light of the known evolutionary behavior of social bots, typical of adversarial settings, which requires constant updates of data and methods~\citep{cresci2021coming}. However, many newly published bot detectors are at least partially based on obsolete training data. 

Annotations of accounts as humans or bots, or any other category, can introduce further bias when the annotation process is opaque or incoherent. As a practical example, cyborgs---accounts that are partly automated and partly managed by humans who can step in as needed to avoid detection---can be labeled as ``humans'' if one only considers their capacity to reply to a message. However, many bot detectors are based on the definition that any account with partial automation is a bot. When evaluating such bot detectors, labeling cyborgs as humans would unfairly inflate the false-positive rate due to the mismatch between the definition adopted by the bot detector versus that of the evaluator. In other cases, the authors change criteria for labeling bots multiple times within the context of the same study, without disclosing any annotation rubric~\citep{gallwitz2022investigating}. First, they label accounts that automatically post news headlines---such as those associated with major newswire agencies---as humans. Then, they ignore suspended accounts or consider them to be human, depending on the analysis. Finally, they label accounts that cross-post tweets through software apps as humans. 

As exemplified above, the use of obsolete or biased datasets with unclear, inconsistent, or shifting definitions and opaque labeling schemes are among the malpractices that affect this field. To avoid bias, account labeling should be performed by multiple independent annotators following a shared and openly accessible rubric, and based on a consistent definition of social bot. Moreover, continuous scores representing the different degrees of ``botness'' (i.e., automation) should be favored in place of binary labels, as the former are better suited to capture nuances in the use of automation.

\begin{center}
    \footnotesize
    \setlength{\tabcolsep}{0.5pt}
    \begin{longtable}{P{0.03\textwidth}P{0.18\textwidth}P{0.25\textwidth}P{0.54\textwidth}}
\toprule
        & \textbf{step} & \textbf{issues} & \textbf{recommendations} \\
        \midrule
        \endfirsthead
\multicolumn{4}{c}{\textit{Table~\ref{tab:checklist} (continued from previous page)}} \\
        \toprule
        & \textbf{step} & \textbf{issues} & \textbf{recommendations} \\
        \midrule
        \endhead
\midrule
        \multicolumn{4}{c}{\textit{Table~\ref{tab:checklist} continues on next page}} \\
        \bottomrule
        \endfoot
\endlastfoot
        \faLightbulbO & study design & contrasting definitions & \checkbox define the characteristics of the targeted bots openly and clearly \\
        &&& \checkbox favor definitions based on objective criteria \\
        &&& \checkbox discuss similarities and differences with other existing definitions \\
        \cmidrule{3-4}
        && inconsistent definitions & \checkbox apply the chosen definition consistently, not on a case-by-case basis \\
        \midrule
        \faFolderOpen & literature review & misrepresentation & \checkbox consider multiple viewpoints \\
        &&& \checkbox present a balanced and nuanced summary of the existing literature \\
        \midrule
        \faDatabase & data collection & data obsolescence & \checkbox favor recent data \\
        &&& \checkbox discuss the potential impact on the analysis of any outdated data \\
        \cmidrule{3-4}
        && data irrelevance & \checkbox ensure that data is relevant to the targeted bots and chosen definition \\
        \cmidrule{3-4}
        && data incompleteness & \checkbox consider including data about multiple types of bots \\
        &&& \checkbox ensure that data is comprehensive and general enough to support the analyses \\
        \cmidrule{3-4}
        && data bias & \checkbox describe data characteristics exhaustively \\
        &&& \checkbox adopt adequate methods to mitigate existing biases \\
        \midrule
        \faTags & data labeling & opaque labeling criteria & \checkbox define labeling criteria openly and clearly \\
        &&& \checkbox ensure consistency between the labeling criteria and the chosen definition \\
        &&& \checkbox describe the labeling process \\
        &&& \checkbox share the labeling rubric \\
        \midrule
        \faCog & model training & inadequate approach & \checkbox ensure consistency between the methodological approach and the goal of the analysis \\
        \cmidrule{3-4}
        && model drift & \checkbox consider the impact of training data on how overfitted or specialized the model is \\
        &&& \checkbox consider training the model with data about multiple types of bots \\ &&& \checkbox consider the recency of training data on the model's capacity to detect current bots \\
        \cmidrule{3-4}
        && information leakage & \checkbox ensure that the model only uses information available at application time \\
        \midrule
        \faSliders & parameter tuning & sensitivity to parameters & \checkbox experiment with multiple bot thresholds and model parameters \\
        \midrule
        \faSearch & model evaluation & unfair comparisons & \checkbox specify inclusion and exclusion criteria for model comparisons openly and clearly \\
        &&& \checkbox favor comparisons with recent, state-of-the-art models \\
        &&& \checkbox include multiple comparisons with similar models: bot definition, approach, $\ldots$ \\
        &&& \checkbox include some comparisons with both strong and weak models \\
        \cmidrule{3-4}
        && overfitting & \checkbox evaluate model against known bots having different characteristics \\ &&& \checkbox evaluate model in-the-wild, by applying it to detect unknown bots \\
        \cmidrule{3-4}
        && sensitivity to parameters & \checkbox carry out sensitivity analyses of bot thresholds and model parameters \\
        \midrule
        \faMagic & model application & model drift & \checkbox consider differences between the training versus the application context \\
        \cmidrule{3-4}
        && inaccurate predictions & \checkbox manually validate a relevant subset of the model predictions \\
        &&& \checkbox ensure validation criteria are consistent with the bot definition used by the model \\
        &&& \checkbox describe the manual validation process \\
        &&& \checkbox consider applying multiple models to confirm predictions \\
        &&& \checkbox consider re-applying the same model multiple times to confirm previous predictions \\
        \midrule
        \faPieChart & analysis & contrasting definitions & \checkbox discuss the implications of the use of alternative bot definitions \\
        \cmidrule{3-4}
        && sensitivity to parameters & \checkbox discuss the implications of the use of alternative bot thresholds and parameters \\
        \cmidrule{3-4}
        && unfair comparisons & \checkbox discuss results in terms of the differences between the evaluated models \\
        \cmidrule{3-4}
        && omitted limitations & \checkbox list limitations openly and comprehensively: data, approach, comparisons, $\ldots$ \\
        &&& \checkbox provide a balanced view of capabilities and shortcomings \\
        \midrule
        \faBullhorn & communication & hyped results & \checkbox avoid sweeping statements \\
        &&& \checkbox consider contrasting viewpoints and results \\
        &&& \checkbox ensure that conclusions are supported by the results \\
        &&& \checkbox adjust language to the breadth and depth of the analysis \\
        \cmidrule{3-4}
        && reproducibility crisis & \checkbox share model training and evaluation code \\
        &&& \checkbox share model training and evaluation data, including manual validation data \\
        &&& \checkbox share model application data: accounts and model predictions \\
        &&& \checkbox discuss code and data sharing limitations \\
        \bottomrule
        \caption{\rev{Summary of the main challenges and issues that may occur at each step of the social bots research process. For each issue, we propose a set of guidelines and practical recommendations. Since not all challenges apply to every study, it is up to the authors of future works to review this table to identify relevant issues specific to their study and follow the corresponding recommendations. The table could also serve as a valuable resource for reviewers and evaluators to assess the rigor of future works in terms of how new works implement the best practices and recommendations listed.}}
        \label{tab:checklist}
    \end{longtable}
\end{center}
\normalsize
 
\subsection*{Practical recommendations}
Table~\ref{tab:checklist} summarizes the main methodological issues that may occur at each step of the social bots research process, and proposes a set of best practices, guidelines, and practical recommendations to mitigate them. 

One significant challenge is the use of contrasting definitions of social bots, which leads to incomparable results and confusion in research. To mitigate this, future studies should openly and clearly define the specific characteristics of the social bots they are targeting. Moreover, future research should adopt definitions consistently, without changing bot criteria and definitions case-by-case. Furthermore, favoring definitions based on objective characteristics---such as the use of automation---over subjective ones, will boost clarity and consistency. When discussing results, researchers should also compare their definitions with existing ones to highlight similarities and differences, also exploring how different definitions could impact the results. A related pressing issue is the use of inconsistent or opaque bot labeling procedures, as these hinder the clarity, comparability, and reproducibility of results. Researchers should document and publish their labeling procedures, including the criteria and tools used to carry out the annotation. Sharing labeled datasets with comprehensive metadata will further enhance transparency, reproducibility, and collaborative efforts.

A further set of recommendations involves the evaluation procedures of social bot detectors. Newly trained bot detectors are often tested on well-known and old bots, which does not accurately reflect their performance in real-world scenarios. Researchers should prioritize using the most up-to-date datasets for their evaluations, as these are likely to better reflect the evolving nature of social bots. Performance assessments should include tests on bots with different and unknown characteristics, to better gauge the detectors’ real-world applicability and to ensure that detectors are robust and effective in dynamic online environments. In addition, attention to validation should not only be the focus of the training and testing phases of a new bot detector, but should also be devoted when using a pre-trained detector developed by others, even if it is well-known and established. A subset of automatically assigned labels and scores should always be manually validated, rather than accepted at face value. This ensures their accuracy and reliability in the given context, which might be different than that on which the detector was developed. As with the aforementioned manual labeling procedures, also this manual validation process should be clearly and openly documented, detailing the criteria and methods used for manual verification of the assigned labels.

Another issue in the evaluation of bot detection performance is the possibility of cherry-picking competitors in such a way to purportedly demonstrate the superior performance of a newly proposed detector. To mitigate this issue, future research should benchmark bot detectors against a comprehensive set of existing detectors, including both strong and weak performers. Moreover, the criteria for the selection of the comparisons should be stated clearly and openly. Sharing evaluation scripts and datasets used in benchmarking will further promote transparency and reproducibility in performance comparisons. In addition, sensitivity analyses should also become a standard practice to address the possible sensitivity of detectors to thresholds and other model parameters, which are often overlooked when reporting bot detection results. These analyses should be reported alongside the main results to provide a clearer picture of the robustness and reliability of the detectors, aiding in the identification of optimal settings and limitations.

Lastly, data availability and documentation are crucial for the progress of bot detection research. The lack of shared and well-documented datasets impedes reproducibility and comparison, and slows down the development of new detectors. Researchers should make their datasets publicly available whenever possible, ensuring they are well-documented with comprehensive metadata, including how the data was collected and labeled. This recommendation does not only involve the datasets used to train or evaluate a detector, such as those labeled manually, but also the datasets on which pre-trained detectors are applied. For the latter, it is crucial to share the data together with the labels or scores automatically assigned by the detectors. If data sharing is constrained by privacy or other issues, researchers should at least provide detailed descriptions of their datasets and the procedures for obtaining them. This practice will support reproducibility and facilitate more rigorous comparisons and advancements in bot detection research.

\begin{figure*}[t]
    \centering
    \begin{subfigure}[t]{.35\textwidth}\includegraphics[width=\textwidth]{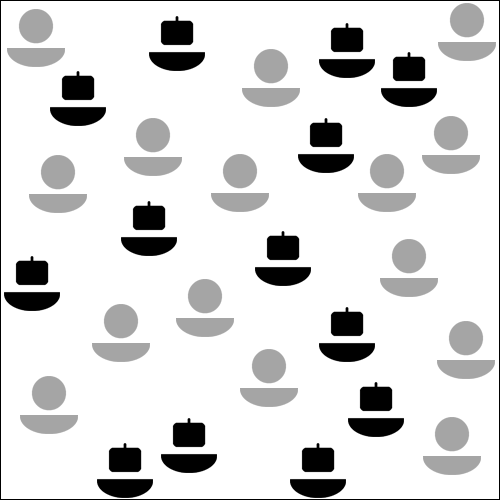}\caption{Simplistic and na\"{i}ve model of social bots (black-colored) and human-operated accounts (grey-colored). Here, all bots are alike, they are evenly distributed, and we have complete knowledge of their numbers and characteristics. According to this model, collecting unbiased and comprehensive bot datasets is possible.}\label{fig:bots-them}\end{subfigure}\hspace{.15\textwidth}\begin{subfigure}[t]{.35\textwidth}\includegraphics[width=\textwidth]{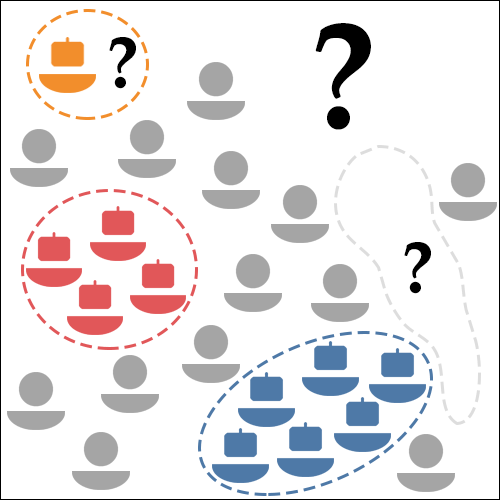}\caption{Realistic model of social bots and human-operated accounts, and of the related knowledge gaps. Here, social bots are organized in botnets, each with peculiar characteristics (color-coded). Among the missing information is the number of bots in known botnets, as well as the number, size, and characteristics of all unknown botnets.}\label{fig:bots-us}\end{subfigure}\caption{The ecology of online accounts: social bots and humans. Differences between na\"{i}ve and realistic models, and their implications for data collection.}
    \label{fig:bot-landscape}
\end{figure*}
 \section{Conceptual issues}
\label{sec:issues-concept}
Beyond methodological challenges, the field is also intricately intertwined with a variety of potentially more subtle---yet profoundly influential---conceptual issues.
These concern how social bots and related phenomena are defined, framed, and understood within the research discourse.
These nuanced challenges, if left unaddressed, have the potential to exert an even more insidious influence on the development of the field and its reception.

\subsection*{Failure to account for context}
In discussing the limitations that currently hinder progress in bot detection, some recent research pointed at flawed data collection practices that fall short of capturing the complexity of the bot space~\citep{hays2023simplistic}. This interpretation of the limitations of the existing datasets implicitly assumes the possibility to encode the full complexity of the bot detection problem space in a dataset. More specifically, this criticism subsumes that building \textit{unbiased} and \textit{comprehensive} bot datasets is possible, and perhaps even convenient. 
Unbiased datasets are needed so as to avoid that peculiar bot and human characteristics leak into the data allowing simple models to achieve good performance on the collected data, while showing poor performance in real scenarios. Furthermore, comprehensive datasets are needed so that expressive models can learn to generalize to all types of existing social bots. Unfortunately, the assumption of the existence of an unbiased and comprehensive dataset is fallacious.

Regarding \textit{bias}, many bot datasets were found to contain biases, in that the accounts therein featured some peculiar characteristics~\citep{hays2023simplistic}. However, this is largely unsurprising since such characteristics are the inevitable consequence of the bots being organized in botnets: groups of accounts created and controlled by a central entity who collectively operates the bots to reach some predefined goal~\citep{zhang2016rise}. Given that all bots belonging to the same botnet are created or operated by the same entity, and pursue the same common goal, they tend to share similarities. When compared to other accounts, the botnet thus appears to have some peculiar characteristics, as highlighted in Figure~\ref{fig:bots-us}. Thus, in most cases, the presence of biases in the existing bot datasets should not be blamed on the creators of the datasets, but rather traced back to the very nature of the social bot phenomenon. 

The assumption about the possibility to build \textit{comprehensive} bot datasets is equally flawed. Creating such datasets would require a uniform and random sample of an adequate number of accounts from the complete distribution of existing bots. This would allow obtaining an accurate representation of the bot landscape that captures the full complexity of the problem space. However, performing a uniform and random sample of the whole population of bots is extremely problematic---if not outright impossible. The task of detecting social bots partly belongs to the fields of information security and open-source intelligence, which are intrinsically characterized by the presence of adversaries who are strongly motivated to remain hidden~\citep{cresci2021coming}. There has always been limited knowledge of the real extent of the bot problem, which makes it hard to track bots on a platform, quantify their numbers, and assess their impact~\citep{varol2022should,mendoza2020bots,tan2023botpercent}. We only have a partial understanding and visibility of the botnets that operate on our online platforms.\footnote{The court dispute between Elon Musk and former Twitter CEO Parag Agrawal on the real number of bots on Twitter is a paramount example of the intrinsic difficulties and ambiguity of tracking social bots.} Figure~\ref{fig:bot-landscape} provides a conceptual view of the landscape of social bots and human-operated accounts: firstly according to a na\"{i}ve interpretation (Figure~\ref{fig:bots-them}) and then in a more realistic representation (Figure~\ref{fig:bots-us}). Figure~\ref{fig:bots-us} highlights the knowledge gaps related to the presence of bots in online platforms. These include \textit{known unknowns}: the exact number of bots belonging to known botnets. But they also include \textit{unknown unknowns}: information about the hidden botnets operating on a platform. How can we create unbiased and comprehensive datasets when we largely ignore the types and numbers of bots populating online platforms? This fundamental question is currently unanswered.

\subsection*{Common misconceptions in social bots research}
\label{sec:misconceptions}
Some recent work posed the idea that the seemingly excellent results in social bot detection could be interpreted as a success story for machine learning and as evidence that the task is now effectively solved~\citep{hays2023simplistic}. According to this perspective, the perceived accuracy and reliability of current detection methods support their suitability for integration into various downstream applications, highlighting the progress made in achieving practical and deployable solutions. While Hays et al.~\citep{hays2023simplistic} pose the question ``Is bot detection a solved problem?'' as a rhetorical device, both their framing and conclusions imply that the field has collectively overlooked critical challenges. This is a problematic foundation. A vast literature has long acknowledged the limitations in social bots research. By sidestepping this body of work, the authors construct a straw man: a field that blindly celebrates near-perfect performance without reflection. Other than boosting the apparent novelty of their critique, this framing misrepresents the field. Unbiased and responsible research in this area should acknowledge the body of literature warning researchers and practitioners about the limitations of bot detection. Below, we report on some notable examples. The first accounts of the difficulties supervised detectors face in detecting sophisticated bots date back to 2013~\citep{yang2013empirical} and intensified in subsequent years~\citep{cresci2017paradigm,grimme2018changing,cresci2019capability}. The limitations of benchmark datasets for bot detection were discussed by Feng et al.~\citep{feng2021twibot}, and the data-related limitations to the generalizability and replicability in social bots research were touched upon by Assenmacher et al.~\citep{assenmacher2021benchmarking}. Several works measured and discussed the generalization deficiencies of bot detectors, with particular emphasis on supervised ones~\citep{echeverri2018lobo,Yang2019botometer,cresci2020decade,sayyadiharikandeh2020detection,yang2020scalable,feng2021twibot}. Others warned about the lack of consistency in bot classifications obtained at different points in time~\citep{rauchfleisch2020false} and with different detectors~\citep{svenaeus2020fantastic,martini2021bot}. Concerns about the accuracy and usefulness of bot detection have also been raised by security and integrity experts at major platforms~\citep{roth2020botornot}, and by independent researchers and the media.\footnote{https://www.nytimes.com/2020/06/16/science/social-media-bots-kazemi.html}\footnote{https://mediawell.ssrc.org/expert-reflections/on-digital-disinformation-and-democratic-myths/}\footnote{https://www.lawfareblog.com/random-toxicity-whats-going-benjaminwittess-mentions} The growing tension between the lasting challenges of bot detection and the increasing technological capabilities of online manipulators, including unprecedented affordances such as those offered by generative AI~\citep{yang2023anatomy,ferrara2023social}, even led some researchers to question the long-term viability of the bot detection task~\citep{boneh2019relevant,grimme2022new}. For the same reasons, some have suggested redirecting part of the ongoing scientific effort away from bot detection and to alternative and more promising tasks, such as detecting information operations and coordinated harmful behavior~\citep{cresci2020decade,roth2020botornot,mannocci2024detection}. This sampling of the existing literature on the challenges and limitations of bot detection paints a very different picture than some recent, optimistic works. This discrepancy introduces a common misconception about social bots research:

\begin{tcolorbox}[sharp corners, colback=white, colframe=black!70, title=Misconception 1]
Social bot detection is a solved task.
\end{tcolorbox}

\noindent The above evidence shows that despite significant efforts devoted for a prolonged time, bot detection is nowhere near to being a solved problem---quite the contrary. 

In framing the bot detection task as a seemingly straightforward problem, some authors also hint at the possibility of easily adapting existing detectors to overcome their current limitations and keep pace with the evolution of more human-like bots~\citep{hays2023simplistic}. 
This example introduces another misconception about bot detection:

\begin{tcolorbox}[sharp corners, colback=white, colframe=black!75, title=Misconception 2]
Bot detection performance can be improved easily.
\end{tcolorbox}

\noindent For many years, we witnessed to a whack-a-mole game between bot developers, with their increasingly sophisticated accounts~\citep{yang2023anatomy,yang2024characteristics}, and bot hunters, equipped with a variety of different detectors~\citep{Yang2019botometer,sayyadiharikandeh2020detection,yang2020scalable}. Looking back at how this arms race has unfolded, we can conclude that none of the technological advances we have experimented with have significantly mitigated the challenges posed by malicious social bots in the long term. It is reasonable to assume that future advances will suffer a similar fate.
Social bots are adversarial, fast-moving targets, characterized by a fast adoption of cutting-edge technology~\citep{cresci2021coming}. Bot detection is therefore an intrinsically challenging task, made even more daunting by the lack of accurate information on the targets of the analysis, the limited collaboration from online platforms, and the rapidly evolving nature of online harms.

Below, we discuss some other misconceptions that undermine the understanding of social bots and of the related literature:

\begin{tcolorbox}[sharp corners, colback=white, colframe=black!75, title=Misconception 3]
All social bots are similar.
\end{tcolorbox}
\begin{tcolorbox}[sharp corners, colback=white, colframe=black!75, title=Misconception 4]
Each bot detector can detect all types of bots.
\end{tcolorbox}

\noindent Both scholars and the general public often assume that a bot detector performing well on some bots will perform equally well on any bot detection task. This assumption is problematic given that there are a plethora of diverse bots, each with its own characteristics~\citep{mazza2022investigating}. Consider, for example, the differences between the bots used to boost the popularity of certain public figures---so-called fake followers~\citep{cresci2015fame}---and those that manipulate trending topics---astroturfers and spammers~\citep{abokhodair2015dissecting}. Or the differences between bots involved in political manipulation~\citep{shao2018spread,caldarelli2020role} versus those aimed at fooling automated trading systems~\citep{tardelli2022detecting,Bello2023}. Some scholars addressed this heterogeneity by designing detectors that aim for generality and broad applicability, such as Botometer, especially in its latest releases~\citep{sayyadiharikandeh2020detection,yang2020scalable}. Others have developed detectors specifically designed to detect certain types of bots. The latter trade generalizability and portability for detection accuracy, and their performance depends heavily on the characteristics of the bots they are designed to detect. For example, a detector designed to detect time-synchronized retweeting actions~\citep{mazza2019rtbust} would likely be useless at detecting mass-following bots~\citep{cresci2015fame}. However, this should not be regarded as a limitation of the bot detector---let alone one deliberately unstated by its developers in order to pass it off as a good product---but rather as an inappropriate use of the detector itself. Limitations in generalizability and portability also affect general-purpose bot detectors, although to a lesser extent than specialized ones. In fact, even general-purpose detectors rely on a limited number of features to estimate whether an account is a bot. Therefore, in general, any detector, specialized or otherwise, has variable performance and its detection capabilities depend on the characteristics of the accounts it is applied to. In conclusion, no single bot detector is capable of detecting all types of bots.

Social bots research is often framed within the broader landscape of mis- and disinformation studies, a field that extends well beyond online manipulation, including critical perspectives on post-truth dynamics, cognitive biases, epistemic trust, ideological polarization, and the structural features of media ecosystems~\citep{spohr2017fake,waisbord2018truth}. The specific relevance of social bots in this broader discourse has largely emerged from concerns over their potential to enable scalable and low-cost manipulation~\citep{shao2018spread,stella2018bots}. These concerns, while widely shared and influential, are themselves being increasingly reexamined, raising questions about the actual impact of bots, the assumptions behind their perceived threat, and the relative emphasis they have received in disinformation research. This surfaces another misconception: 

\begin{tcolorbox}[sharp corners, colback=white, colframe=black!75, title=Misconception 5]
Social bots are mainly responsible for the spread of disinformation.
\end{tcolorbox}

\noindent An unbiased analysis of the existing literature paints a dubious picture of the role of social bots. For example, while some studies have concluded that bots play a prominent role in the spread of problematic content~\citep{shao2018spread,stella2018bots}, others have reached the opposite conclusion~\citep{vosoughi2018spread,gonzalez2021bots,seckin2024mechanisms}. 
The existing literature has focused almost exclusively on detecting bots and characterizing their behavior, leaving the fundamental task of measuring the impact of bot malfeasance largely unexplored~\citep{cresci2020decade}. For these reasons, we currently lack scientific consensus and conclusive evidence about the role of social bots and their effectiveness in influencing online users. What we do know, however, is that bots are only one of many agents involved in the spread of mis- and disinformation~\citep{starbird2019disinformation,roth2020botornot}. Examples of other agents are state-sponsored trolls, users who collude and coordinate for malicious purposes, superspreaders, and even willing but unwitting individuals~\citep{starbird2019disinformation,DeVerna2022FIB}. Each of these agents represents a potential threat to safe and trusted online platforms, and a thriving area of research and experimentation. It is therefore critical to balance efforts in all of these directions, avoiding the pitfall of overstudying some while overlooking others, based on unsupported decisions.

The previous misconceptions might induce readers to think that social bot research has led to flawed---if not outright useless---results. Such claims have been recently made by Hays et al.~\citep{hays2023simplistic} and Gallwitz \& Kreil~\citep{gallwitz2022investigating}---the latter even in sensationalist terms.
These claims lead to our last, but by no means least important, misconception:
\begin{tcolorbox}[sharp corners, colback=white, colframe=black!75, title=Misconception 6]
All social bots research is flawed and bot detection results are useless.
\end{tcolorbox}

\noindent We see at least three strong arguments against this thesis. First, despite the limitations of bot detectors, there are several glaring examples of bot studies that were able to bring to light demonstrably harmful campaigns. For example, detectors developed as part of some scientific efforts were later deployed on online platforms and used to remove large numbers of malicious accounts~\citep{yang2014uncovering}. Similarly, the results of some studies led platforms to remove accounts identified as malicious bots~\citep{ferrara2022twitter,yang2023anatomy}. In several other cases, scientific findings about bot activity were later found to be consistent with independent platform removals of malicious accounts~\citep{nizzoli2020charting,tardelli2022detecting}, confirming the accuracy of the scientific findings. These cases represent just some of the success stories of social bot research. Therefore, even if no universal bot detector exists, and despite the many caveats to consider in bot detection, being able to detect \textit{some} malicious bots puts us in a more advantageous position than being able to detect \textit{none}. 

Second, the methodological rigor and practical usefulness of some social bots studies are evidenced by corrective actions taken by government agencies based on academic research. For instance, in November 2014, the U.S. Securities and Exchange Commission (SEC) issued an alert to raise awareness about stock market manipulations exacerbated by social bots,\footnote{Updated Investor Alert: Social Media and Investing -- Avoiding Fraud (Nov. 12, 2014) \url{https://www.sec.gov/resources-for-investors/investor-alerts-bulletins/ia_socialmediafraud}} following the findings of earlier studies~\citep{10.1145/2090150.2090161}. Subsequent research confirmed these suspicions on multiple occasions~\citep{cresci2019cashtag,nizzoli2020charting}, contributing to calls for greater government oversight of social bots~\citep{gorwa2020unpacking,yan2023} and discussions on regulating bot freedom of expression~\citep{lamo2019}. Both the 2018 and 2022 EU Code of Practice on Disinformation, as well as investigations by the U.S. Congress and the UK Parliament’s Digital, Culture, Media and Sport Committee, have also been informed by research on social bots. Furthermore, the concern over bot infiltration was highlighted when Elon Musk initially decided to abandon his purchase of Twitter, citing misleading information about the platform's financial health and bot prevalence~\citep{varol2022should}---a claim also supported by Peiter Zatko, Twitter’s former head of security, who exposed serious deficiencies in the company's security.\footnote{What the Twitter Whistleblower Disclosure Means for Elon Musk: \url{https://time.com/6207992/twitter-bots-elon-musk/}} 

Third, the benefits of social bot research extend beyond the detection of malicious bots. For example, research and experimentation on social bots led to the development of neutral bots used to assess the level of political polarization and bias on a platform~\citep{chen2021neutral}; ``news bots'' used for journalistic purposes to curate, aggregate, and distribute content gathered from multiple sources~\citep{lokot2016news}; and even bots used for content moderation~\citep{bilewicz2021artificial,askari2024incentivizing}. In addition, social bot research has contributed to the early development of other neighboring fields. The early research on social bots, which dates back to 2010~\citep{cresci2020decade}, provided an important foundation, allowing the field to draw upon several years of accrued experience when widespread concerns regarding misinformation, state-backed trolls, and coordinated inauthentic behavior surged in 2016 and subsequent years. In other words, early results on detecting and characterizing social bots informed strategies for detecting and mitigating other related forms of online manipulation. In light of these considerations, research on social bots---imperfect as it is---seems far from useless. In fact, social bots research perfectly exemplifies the process leading the general advancement of science: the accumulation of knowledge resulting from certain research efforts, other than solving local problems, also fertilizes the broader scientific ecosystem, providing data, methodologies, tools, and insights for further scientific advancements in close---and sometimes not so close---fields.

 \section{Challenges in the post-API era}
\label{sec:api}
In addition to the widespread biases and misconceptions detailed above, the field of social bot research confronts significant obstacles due to changes in social media platform policies. In the API era, most social bot studies have focused on Twitter/X due to its free API and the ease of large-scale data acquisition. This hyperfocus on X has led to the neglect of other platforms, creating a bias in social bots research that largely ignores other platforms. However, this data policy ceased in 2023 when X terminated free data access for researchers.\footnote{https://www.theverge.com/2023/5/31/23739084/twitter-elon-musk-api-policy-chilling-academic-research} Similarly, Reddit restricted free data access, and Meta announced the discontinuation of CrowdTangle, a crucial tool for researchers studying Facebook and Instagram.\footnote{https://www.niemanlab.org/2024/03/a-window-into-facebook-closes-as-meta-sets-a-date-to-shut-down-crowdtangle}

The lack of access to fresh social media data severely hampers researchers' ability to monitor bot activities and assess their influence. It significantly impedes the collection of new bot samples necessary for studying their characteristics and training novel machine learning classifiers. Even if researchers can develop new classifiers, deploying them at scale is challenging, exposing social media users to potential manipulation. Bot operators, on the other hand, remain largely unaffected. Leveraging burner and virtual cellphones, they can manage bot accounts across platforms directly, bypassing API restrictions. The advent of AI-powered social bots, which evade current bot detection models~\citep{yang2023anatomy,yang2024characteristics}, further emphasizes the importance of data availability in support of social bot research.

Despite these setbacks, Europe's Digital Services Act (DSA) has offered some hope by mandating large social media platforms to grant researchers data access upon reasonable requests.\footnote{https://digital-strategy.ec.europa.eu/en/policies/digital-services-act-package} Platforms like TikTok, Meta, and Google have initiated new data access programs. However, these programs are significantly limited compared to previous access levels.  Their opaque and stringent application review processes also leave their efficacy in question~\citep{jaursch2024dsa}. Besides traditional social media, the rise of decentralized platforms such as Mastodon and BlueSky introduces new dynamics. Their openness facilitates data access for researchers but also exposes them to exploitation by malicious actors. The decentralized nature of these platforms complicates efforts to combat threats such as malicious social bots, presenting novel challenges to both users and the research community. Moreover, the limited number of participants, which has not yet reached a critical mass, and the difficulty in creating a reliable ground truth, further contribute to hindering bot detection research on these platforms. Overall, the new data accessibility landscape and the change in the underlying technologies, represent an opportunity and a call for further research in social bot detection.
 \section{A call for moral responsibility}
\label{sec:discussion}
Our analysis of the drawbacks and limitations of recent studies brought to light the need for responsibility when discussing results about social bots. The way in which new findings are presented in this and in neighboring fields can influence not only the next iterations of research, but also industry practices, policymaking, and public opinion. To this end, it is paramount to avoid repeating the mistakes and propagating the common misconceptions that currently haunt the social bot literature and that fuel ambiguities, generalized misunderstandings, and friction among scholars. 

The everyday challenges that we face as researchers in the broad area of misinformation are an accurate reflection of those faced by our society. As authors, reviewers, and readers of new research in this field, we have the moral obligation of refraining from falling for the same biases, and exacerbating the same issues, that we frequently encounter in our analyses. 
Misinformation often surfaces as «an accurate fact set in a misleading context»~\citep{starbird2019disinformation}. Perpetuating the methodological and conceptual issues addressed in this paper contributes to creating flawed or unreliable research. Cherry-picking data, references, claims, or results from a cited article are examples of misleading context in which misinformation thrives~\citep{west2021misinformation}. Similarly, while making hyped or sensationalist claims can help publish a paper and accrue citations, such claims also contribute to creating misleading contexts and unrealistic expectations, as well as to increasing scientific tribalism. 

Making tangible progress in the science of misinformation requires that we address many conceptual, practical, and ethical challenges. Doing so involves embracing the intrinsic complexity of the phenomenon, considering multiple viewpoints, and providing nuance rather than naivety. Sweeping statements such as ``the creators of bot datasets are responsible for the failure of the field,'' ``all social bots research is flawed or useless,'' or even ``this new bot detector has flawless performance'' do not contribute to achieving these goals.

In conclusion, should we fail to respect our moral obligation, we would produce \textit{biased and unreliable research}, further worsening the problems that currently undermine the credibility of our field~\citep{west2021misinformation,altay2023misinformation}. In a 2019 piece published in \textit{Science}, Derek Ruths compared contrasting findings about social bots, commenting that «research on misinformation has come to resemble the very thing it studies»~\citep{ruths2019misinformation}. It is our moral responsibility to reverse this fatal course. This can only be achieved via responsible research that fosters nuanced, unbiased, and balanced viewpoints, and by a review process that adheres to the same principles. This article aims to make a contribution in this direction by debunking common fallacious arguments adopted by both proponents and opponents of social bots research, as well as providing directions toward sound methodologies for future research in the field.

 \section{Acknowledgments}
This work was partially supported by the European Union -- Next Generation EU, Mission 4 Component 1, for project PIANO (CUP B53D23013290006) and for the ERC project DEDUCE under grant \#101113826.; by project \texttt{DESIRE 2.0} (DissEmination of ScIentific REsults) funded by IIT-CNR; by the Volkswagen Foundation as part of the \texttt{Bots Building Bridges} project, under the ``Artificial Intelligence and the Society of the Future'' initiative; by DARPA (awards W911NF-17-C-0094 and HR001121C0169); and by the Knight Foundation.
 
\pagebreak

\bibliographystyle{plainnat}
\bibliography{references}

\end{document}